\PassOptionsToPackage{table}{xcolor} 
\documentclass[sigconf]{acmart}
\usepackage{fancyhdr}
\usepackage[english]{babel}
\usepackage{blindtext}
\usepackage{amsmath}
\usepackage{enumitem}
\usepackage{algorithm}
\usepackage{algpseudocode}
\usepackage{tabularx}
\usepackage{booktabs}
\usepackage{caption}
\usepackage{graphicx}
\usepackage{textcomp}
\usepackage{xspace}

\usepackage{filecontents}
\usepackage[english]{babel}
\usepackage{blindtext}

\usepackage{placeins}

\usepackage{tikz}

\usepackage{subfigure}
\usepackage{float}
\usepackage{multirow}
\usepackage{multicol}

\usepackage{circledsteps}

\usepackage{color}

\usepackage{footnote}
\usepackage{array}

\usepackage{enumitem}
\setlist{nosep}

\usepackage{diagbox}

\usepackage{url}

\usepackage{booktabs}
\usepackage{makecell}

\usepackage[hang,flushmargin]{footmisc}

\usepackage{pifont}  
\usepackage{array}   

\definecolor{codegreen}{rgb}{0,0.6,0}

\renewcommand\footnotetextcopyrightpermission[1]{} 
\setcopyright{none}

\acmDOI{}

\acmISBN{}

\acmPrice{}

\begin{document}
\newcommand{\TIDP}{\textsf{Pegasus}\xspace}
\newcommand{\eg}{e.g.,\xspace}
\newcommand{\ie}{i.e.,\xspace}

\title{\TIDP: A Universal Framework for Scalable Deep Learning Inference on the Dataplane}


\author{Yinchao Zhang\textsuperscript{$\star$}
\hspace{8pt}
Su Yao\textsuperscript{$\star$}
\hspace{8pt}
Yong Feng\textsuperscript{$\star$}
\hspace{8pt}
Kang Chen\textsuperscript{$\dagger$}\textsuperscript{$\star$}
\hspace{8pt}
Tong Li\textsuperscript{$\spadesuit$}
\hspace{8pt}
Zhuotao Liu\textsuperscript{$\star$}
\\
Yi Zhao\textsuperscript{$\diamond$}
\hspace{8pt}
Lexuan Zhang\textsuperscript{$\star$}
\hspace{8pt}
Xiangyu Gao\textsuperscript{$\star$}
\hspace{8pt}
Feng Xiong\textsuperscript{$\S$}
\hspace{8pt}
Qi Li\textsuperscript{$\star$}
\hspace{8pt}
Ke Xu\textsuperscript{$\star$}}

\affiliation{
\institution{
  \textsuperscript{$\star$}Tsinghua University
  \hspace{8pt}
  \textsuperscript{$\spadesuit$}Renmin University of China
  \hspace{8pt}
  \textsuperscript{$\S$}Beihang University
  \\
  \textsuperscript{$\diamond$}Beijing Institute of Technology
  \hspace{8pt}
  \textsuperscript{$\dagger$}Beijing National Research Center for Information Science and Technology
  \city{}
  \country{}
}
}

\renewcommand{\shortauthors}{X.et al.}

\begin{abstract}

The paradigm of Intelligent DataPlane (IDP) embeds 
deep learning (DL) models on  the network dataplane to enable 
intelligent traffic analysis at line-speed.
However, the current use of the match-action table (MAT) abstraction on the dataplane is misaligned with DL inference, leading to several key limitations, including accuracy degradation, limited scale, and lack of generality.
This paper proposes \TIDP to address these limitations. 
\TIDP translates DL operations into three dataplane-oriented primitives to achieve generality: \textsf{Partition}, \textsf{Map}, and \textsf{SumReduce}. 
Specifically, \textsf{Partition} “divides” high-dimensional features into multiple low-dimensional vectors, making them more suitable for the dataplane; \textsf{Map} "conquers" computations on the low-dimensional vectors in parallel with the technique of \textit{fuzzy matching}, while \textsf{SumReduce} "combines" the computation results.
Additionally, \TIDP employs \textit{Primitive Fusion} to merge computations, improving scalability.
Finally, \TIDP adopts full-precision weights with fixed-point activations to improve accuracy.
Our implementation on a P4 switch demonstrates that \TIDP can effectively support various types of DL models, including Multi-Layer Perceptron (MLP), Recurrent Neural
Network (RNN), Convolutional Neural Network (CNN), and AutoEncoder models on the dataplane. 
Meanwhile, \TIDP outperforms state-of-the-art approaches with an average accuracy improvement of up to 22.8\%, along with up to 248x larger model size and 212x larger input scale.

\end{abstract}

\maketitle

\section{Introduction}\label{intro}

In recent years, there has been a growing demand for Intelligent DataPlane (IDP), which leverages data-driven learning models to overcome the limitations of traditional rule-based approaches~\cite{taurus, netbeacon, leo, holterbach2019blink, yin2022practical}.
By utilizing high-performance programmable hardware~\cite{p4, yang2022trio}, IDP supports forwarding-native execution of learning models, enabling intelligent traffic analysis at line-speed, without affecting network 
throughput and latency.

The core challenge in realizing IDP lies in the fact that switch dataplanes are primarily optimized for high-speed packet processing using the match-action table (MAT) abstraction~\cite{mat}, which inherently limits their ability to represent  learning models. 
While many recent works \cite{leo,netbeacon,zheng2022iisy} have explored tree-based models due to the similarity between their decision processes and the MAT abstraction, there remain scenarios that demand more expressive and versatile models. Consequently, the community has also explored to 
incorporating neural network (NN)-based models in IDP. 
However, the MAT abstraction on the dataplane lacks the flexibility to support complex computations such as multiplication and exponentiation, which are essential for DL. 

To address this challenge, 
two main approaches have been proposed: computation simplification~\cite{innetwork_nn, automating_innetwork_ml, qin2020line} and computation bypassing~\cite{xiong2019switches}.
Computation simplification simplifies operations, for example,  by binarizing the entire model. For instance, N3IC~\cite{n3ic} replaces multiplication with binary XNOR and population count (popcnt) operations, directly implementing binary Multi-Layer Perceptron (MLP) within the MAT on SmartNICs. Computation bypassing avoids computation by storing input-output relationships on the dataplane, recording an enumerative mapping from input bit strings to output bit strings, as demonstrated in BoS~\cite{bos}.

However, both methodologies suffer from three key limitations: 

\noindent \textbf{Accuracy.} Accuracy refers to how well a model accomplishes its task, measured by metrics such as precision, recall, or F1-score, depending on the specific traffic analysis tasks. 
Computation simplification, such as model binarization in N3IC~\cite{n3ic}, degrades accuracy due to the reduced numerical range. 
For example, N3IC may lead to accuracy degradation in VPN traffic classification tasks~\cite{bos}. In contrast, computation bypassing through mapping can improve accuracy compared to computation simplification.
However, the mapping has limitations in scale (see below). This limitation forces reductions in input precision or dimensionality, leading to a loss of critical information necessary for accurate predictions~\cite{van2009dimensionality, zhou2016low_bitwidth, hubara2018quantized}.

\noindent \textbf{Scalability.}
Scalability represents the ability to perform DL inference at a larger scale, and this larger scale applies to both the input scale and the model size.
Computation simplification encounters difficulties in both the input scale and the model size. N3IC~\cite{n3ic} cannot scale its throughput on the dataplane, such as in Barefoot Tofino architecture switches~\cite{tofino}, due to additional limitations on binary operations in high-speed environments (e.g., shorter processing cycles allowing only one binary operation on one variable per MAT stage). Computation bypassing methods like BoS~\cite{bos} suffer from limited input scale (e.g., a 21-bit input requires $2^{21}$ table entries, exceeding the capacity of the Barefoot Tofino 2 programmable switch~\cite{tofino2}), resulting in poor scalability.

\noindent \textbf{Generality.} Generality refers to whether the system can support different DL operations, and thereby utilize these operations to perform inference for various models. The computation simplification strategy of N3IC is limited to Matrix Multiplication (MatMul) and fails to generalize to other DL layers, such as Batch Normalization (BN) and activation functions. BoS~\cite{bos} also faces the generality problem. It processes a small number of inputs per time step for binary Recurrent Neural Network (RNN), making it unsuitable for other model types. This limitation conflicts with the current trend in networking to design specialized models for different tasks and to leverage larger input scales to capture more complex relationships~\cite{li2024m3_trend, wang2024nn_trend, wu2024netllm_trend, gallo2020real_trend, alqiam2024transferable_trend}. The generalization issue restricts flexibility and limits the broader potential of IDP.

\begin{table}[tb]
\footnotesize
\begin{tabular}{lccccc}
\hline
 Design& \ding{182} & \ding{183} & \ding{184} & \ding{185} & \ding{186} \\ \hline
Accuracy             &            &            & \ding{51}  &            &            \\
Scalability          &            & \ding{51}  &            & \ding{51}  & \ding{51}  \\
Generality           & \ding{51}  &            &            &            &            \\ \hline
\end{tabular}
\caption{The goals of different designs in Pegasus.}
\vspace{-0.5cm}
\label{tab:introduction}
\end{table}

This paper proposes \TIDP to address the three limitations above with five tightly-coupled designs. 
\ding{182}~For a wide range of model types, \TIDP translates operations (e.g., MatMul, BN and ReLU) within DL layers into three primitives: \textsf{Partition}, \textsf{Map}, and \textsf{SumReduce}. Specifically, \ding{183}~\TIDP uses \textsf{Partition} to divide one-time operations on the entire input into multiple fine-grained computations on minimal input units, uses \textsf{Map} to retrieve precomputed results from mapping tables for each unit and uses \textsf{SumReduce} for aggregation to handle multi-input scenarios, which reduces table sizes.
This method reduces the number of inputs that each table has to process.
\ding{184}~Therefore, \TIDP is able to adopt full-precision model weights (precomputed with full-precision parameters) while using fixed-point numbers for activation representations instead of binary numbers (since an 8-bit number query requires only $2^8$ table entries). This leverages a wider numerical range, allowing the capture of more detailed and precise information crucial for accurate model predictions.
\TIDP also adopts two additional methods to further optimize the primitive implementation.
\ding{185}~Divide the input into minimal units increases the number of lookups, placing considerable additional pressure on memory access bandwidth. To mitigate this overhead, we introduce \textit{fuzzy matching}, which groups multiple units together and maps it to a corresponding table entry, enabling a single lookup to cover multiple units, effectively reducing memory access bandwidth consumption.
\ding{186}~Additionally, \TIDP adopts \textit{Primitive Fusion} to merge multiple operations, thereby reducing the total number of tables.
Table \ref{tab:introduction} shows how these designs contribute to addressing the three limitations of previous approaches.

\begin{table}[tb]
\footnotesize
\begin{tabular}{cccc}
\hline
Prior Works & Accuracy & \begin{tabular}[c]{@{}c@{}}Model\\ size\end{tabular} & \begin{tabular}[c]{@{}c@{}}Input\\ scale\end{tabular} \\ \hline
N3IC \cite{n3ic} (binary MLP)  & 22.8\% $\uparrow$ & 248x $\uparrow$ & 29x $\uparrow$  \\
BoS \cite{bos} (binary RNN)    & 17.9\% $\uparrow$ & 237x $\uparrow$ & 212x $\uparrow$ \\
Leo \cite{leo} (Decision Tree) & 17.2\% $\uparrow$ & -               & -               \\ \hline
\end{tabular}
\caption{\TIDP v.s. Prior Works.}
\vspace{-0.5cm}
\label{tab:comparison}
\end{table}

\textbf{Contributions}.
The major contribution of this paper is the design, implementation and evaluation of \TIDP, the first IDP design that supports multiple DL models on commodity programmable switches. Our implementation on the P4 switch demonstrates that \TIDP can effectively support various model types on the dataplane, including MLP, RNN, Convolutional Neural Network (CNN), and AutoEncoder. Experiments show that \TIDP can support 3840 bit input scale, 6083 Kb model size, and achieves an average classification accuracy of 97.3\% (Table~\ref{tab:comparison} gives a preview of \TIDP's benefits, see Table \ref{tab:performance} for the full results).

\section{Background and Motivation}\label{back}


\noindent \textbf{Deep Learning.}
Deep Learning (DL) utilizes neural networks composed of multiple layers to model complex patterns in data~\cite{kelleher2019deep}. 
The implementation of DL involves a variety of operations that process data through different types of layers, such as fully connected (FC), convolutional (Conv), activation (Act), normalization (Norm), pooling (Pool), recurrent (Rec), and embedding (Emb) layers.

In DL, each layer performs specific mathematical operations. For instance, FC layers compute weighted sums of inputs plus biases, enabling the network to capture linear relationships. Conv layers apply convolution operations to detect local patterns like edges in images. Rec layers handle sequential data by maintaining a hidden state that captures temporal dependencies. Activation functions like ReLU, Softmax, and tanh introduce nonlinearity, allowing the network to learn complex, non-linear relationships.
Norm layers adjust the input distributions to subsequent layers, enhancing model stability and performance.
Pooling layers reduce the spatial dimensions of data, decreasing computational load and controlling overfitting by summarizing features. Embedding layers transform discrete data into continuous vector spaces, which is particularly useful for capturing temporal features in time series data.
All these operations involve intensive computations, such as Matrix Multiplications (MatMul), convolutions, and non-linear transformations.

\noindent \textbf{Programmable Dataplane.}
The emerging programmable switches~\cite{mat, yang2022trio, shrivastav2022programmable} offer flexible dataplane programmability, allowing developers to execute custom processing logic on each data packet.
Many programmable switches today can be programmed using the P4~\cite{p4} language, a domain-specific language based on the match-action table (MAT)~\cite{mat} abstraction.
The MAT abstraction extracts fields from packet headers and matches them against flow tables, where matched entries specify the actions to be executed on the packets. 
While the MAT abstraction provides significant flexibility for designing network functions, its practical implementations often face critical limitations.

For instance, the Protocol-Independent Switch Architecture (PISA)—one of the most widely adopted implementations—supports only basic integer operations such as bitwise operations (e.g., NOR, XNOR), shifts, addition, and subtraction. It does not support floating-point numbers, multiplication, division, nor exponential operations—operations that are essential for DL inference computations.
Secondly, the resources available for MAT on the dataplane are limited. 
For instance, on Barefoot Tofino 2, each pipeline only has 20 MAT stages, with each stage equipped with 10 Mb of SRAM, 0.5 Mb of TCAM, and a 1024-bit-wide Action Data Bus~\cite{tofino2}. 
Given that DL involves numerous operations across multiple layers, the 20 MAT stages and 1024-bit bus make it difficult to meet the computational and data transfer demands.

\noindent \textbf{Why DL on the dataplane?}
The increasing demand for real-time, intelligent network traffic analysis has created a need to deploy learning models directly on the dataplane switch~\cite{seufert2024marina, perry2023deep}, enabling tasks like Intrusion Prevention systems (IPS) to analyze and block malicious traffic with terabit throughput and nanosecond-level latency.
Traditional approaches~\cite{leo,netbeacon,zheng2022iisy} often rely on tree-based models, which are valued for their simplicity and interpretability. DL complement these methods by offering significant advantages in addressing certain unique challenges of networking:
(1)~The transmission of network data inherently exhibits temporal characteristics, and DL models, such as RNNs and 1D CNNs, are well-suited to capture temporal patterns, making them better fit network-specific tasks. (2)~DL can extract features directly from raw packets, overcoming the difficulties of complex feature computation in the constrained dataplane environment~\cite{bos}. (3)~The networking field often lacks labeled data~\cite{zhou2024trafficformer} and needs to address continually emerging new attacks~\cite{mirsky2018kitsune}, such as zero-day attacks~\cite{han2023anomaly,yang2022griffin}, making the unsupervised learning capabilities of DL an invaluable tool for adapting to these dynamic and unpredictable scenarios.

\noindent \textbf{Motivation.}
DL inference typically involves highly compute intensive operations, which conflict with the flow table-centric dataplane. 
This requires developers to design DL inference implementations that better align with the dataplane MAT abstraction.

N3IC uses XNOR and population count (popcnt, counting the number of 1s in the binary representation) to replace the multiplication and addition operations in MatMul. This enables the implementation of a simple binary Multi-Layer Perceptron (MLP) on the computation-constrained dataplane.
However, binarizing the entire model reduces precision, leading to accuracy degradation.
Moreover, this method does not support other DL operations, such as activation functions, limiting its generality.
Finally, this approach has poor scalability, making it hard to fit within switch pipelines. 
For example, a 128-bit to 64-bit MatMul requires 64 XNOR and popcnt operations, with each popcnt taking up 14 switch stages~\cite{bos}. 

In contrast, state-of-the-art BoS~\cite{bos} bypasses all DL operations by looking up the mapping from input bit strings and output bit strings. 
This approach allows binary activations only at the input and output layers, while enabling full-precision computation within the model. 
This improves accuracy to some extent compared to N3IC.
However, this method limits input scalability, requiring $2^n$ entries for an n-bit input, resulting in poor overall scalability. 
This limitation brings two additional issues. 
First, input binarization is required to increase dimensionality and boost accuracy.
This binarization, however, reduces the numerical range of input data, leading to accuracy degradation, as evidenced by our experiments in \S \ref{subsec:performance}. 
Second, the restricted input scale limits the method’s generality, making it primarily suitable for models like Recurrent Neural Networks (RNN), where small inputs are processed at each time step.

We noticed that a recent work, Taurus~\cite{taurus}, explores the design of a novel ASIC by incorporating additional hardware resources to enable DL inference. 
In this paper, We focus exclusively on implementing DL inference on commodity programmable switches.

\section{Design Overview}\label{over} 

\subsection{Design Goals}

We propose \TIDP to achieve higher accuracy, greater scalability, and generality in supporting various DL models.
(1)~\TIDP introduces three primitives, including \textsf{Partition}, \textsf{Map}, and \textsf{SumReduce}, to decompose DL models into a sequence of primitives, achieving generality. (2)~\TIDP uses \textsf{Partition} to divide input into segments, uses \textsf{Map} with \textit{fuzzy matching} to retrieve precomputed results for each segment, and applies \textsf{SumReduce} to aggregate results through summation. 
Additionally, \TIDP employs \textit{Primitive Fusion} to merge multiple primitives, reducing the number of operations. These methods enable \TIDP to efficiently handle larger model scales, achieving scalability.
(3)~Finally, \TIDP employs full-precision weights and fix-point activations to enhance model accuracy.

\subsection{\TIDP Architecture}

\begin{figure}[tb]
	\centering
	\includegraphics[width=0.45\textwidth]{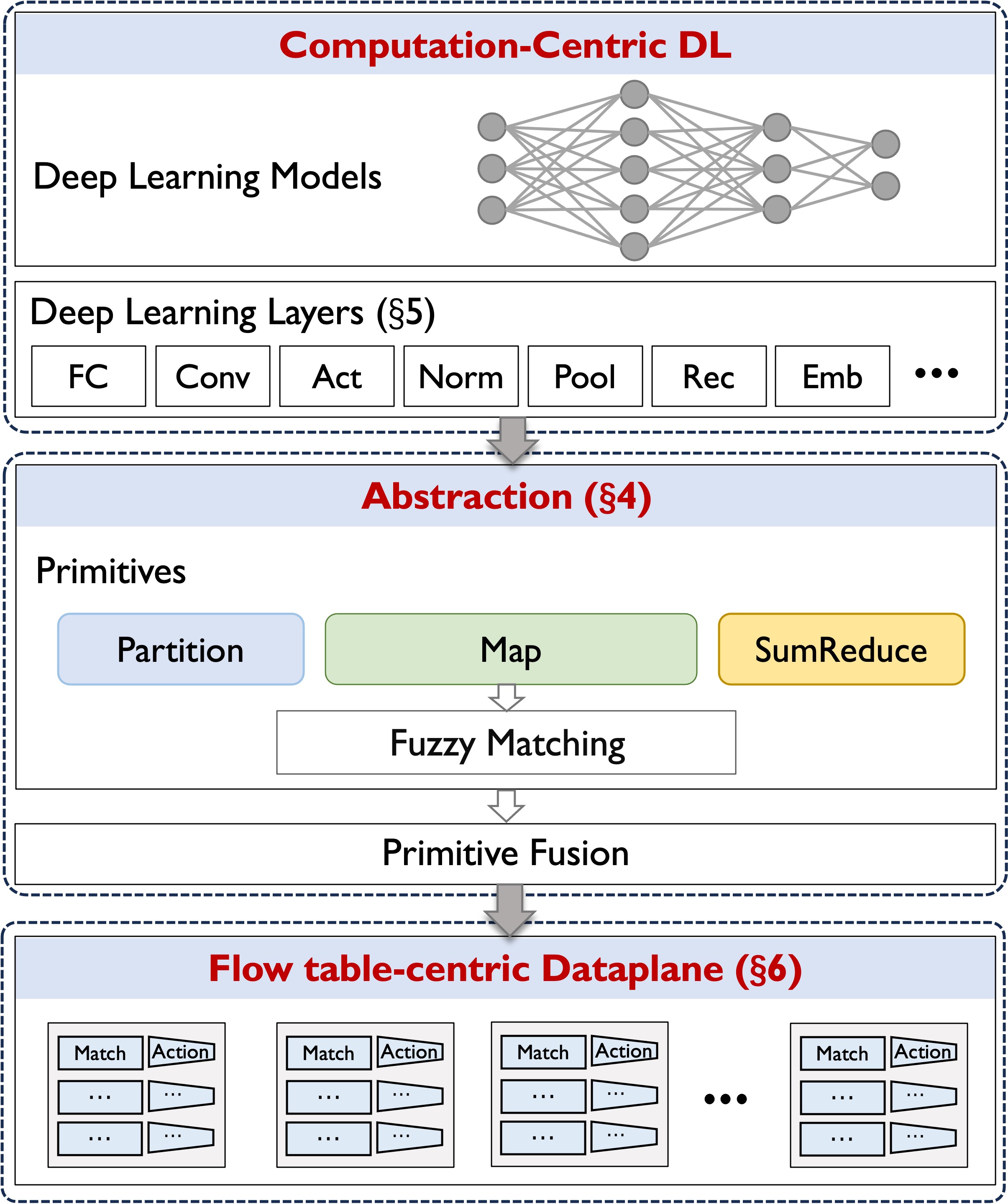} 
	\caption{\TIDP Architecture.}
	\label{fig:overview}
         \vspace{-0.4cm}
\end{figure}

Figure \ref{fig:overview} shows the hierarchical design of \TIDP. 
DL layers are composed of various DL operators, which are further converted into primitives for computation.
The design of these primitives is dataplane-oriented and can be integrated tightly with the MAT abstraction. The lowest-level implementation of the primitives needs to satisfy the limitations of the specific programmable switch.

We analyzed common operations in DL (detailed in \S \ref{DLO}) and found that many functionalities can be realized through parallel data operations, necessitating the design of the \textsf{Map} primitives. DL often requires simple sum aggregation, and implementing this process on the dataplane is not complex. The \textsf{SumReduce} primitives are proposed.
Finally, to enable data flow between primitives, the \textsf{Partition} primitive is needed.
DL operators are then represented using these primitives.
For example, consider a MatMul operation. We can use \textsf{Partition} to divide the input, apply \textsf{Map} to compute the product of each segment with the target matrix, and obtain the final result through \textsf{SumReduce}.
\section{\TIDP Primitives}
\label{sec:primitives}

\subsection{Primitives}\label{pri}
\TIDP primitives fall into three categories: \textsf{Partition}, \textsf{Map}, and \textsf{SumReduce}, as illustrated in Table \ref{tab:primitives}. 
The three primitives can be combined in varying quantities and orders to assemble various DL operators, enabling the construction of distinct DL models.
Specifically, as the dataplane is better suited for handling multiple parallel small-scale computations rather than a single large-scale operation, the \textsf{Partition} primitives divide the multi-dimensional input vector into sub-vectors, reducing computational complexity.
\textsf{Map} primitives execute specific functions (e.g., activation function and batch normalization) on each segment of inputs. 
Fuzzy matching (\S\ref{subsec:clustered_mapping_table}) efficiently supports multiple \textsf{Map} primitives with minimal storage resources and table lookups.
\textsf{SumReduce} primitives perform element-wise summation on multiple vectors, resulting in an aggregated vector.
These primitives are simple enough to be implemented using the MAT abstraction.
More importantly, \TIDP employs primitive fusion (\S\ref{subsec:primitive_fusion}) to reduce resource overhead and improving the scalability.

\begin{table}[tb]
\centering
\begin{tabular}{l|l}
\hline
\rowcolor[HTML]{EFEFEF} 
Primitives & Expression                                                  \\ \hline
\textsf{Partition}  & $\textsf{Partition}(X) = \{X_1, X_2, \dots, X_k\}$                   \\ \hline
\textsf{Map} & \begin{tabular}[c]{@{}l@{}}$\textsf{Map}(\mathcal{F}, \{X_1, X_2, \dots, X_k\}) = $\\ $\{F_1(X_1), F_2(X_2), \dots, F_k(X_k)\}$\end{tabular} \\ \hline
\textsf{SumReduce}  & $\textsf{SumReduce}(\{X_1, X_2, \dots, X_k\}) = \sum_{i=1}^{k}{X_i}$ \\ \hline
\end{tabular}
\caption{Primitives in \TIDP. $X$ is the input vector, $X_i$ represents the $i$-th segment of $X$, and $\mathcal{F}$ is a set of functions including $F_i$.}
\vspace{-0.5cm}
\label{tab:primitives}
\end{table}

\subsection{Fuzzy Matching}
\label{subsec:clustered_mapping_table}

Instead of retrieving results through exhaustive and non-scalable input-output mapping table lookups, fuzzy matching groups multiple input units into a vector and executes a feature-threshold-based search on the vector.

\begin{figure}[tb]
	\centering
	\includegraphics[width=0.47\textwidth]{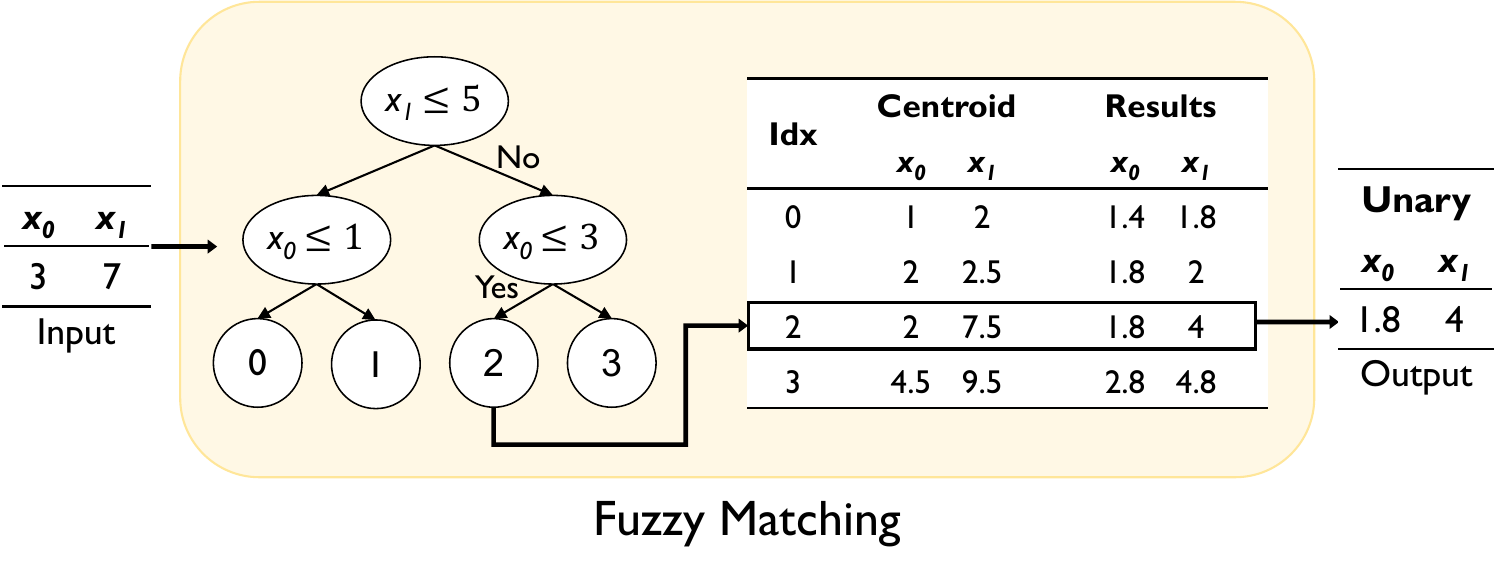}
	\caption{Implementation of a \textsf{Map} primitive: how input sub-vector $(3,7)$ retrieves results $(1.8,4)$ as the approximation of $f(X_i) = 0.4X_i + 1$.}
	\label{fig:cache_search}
         \vspace{-0.3cm}
\end{figure}

\noindent \textbf{Fuzzy Indexing.}
Specifically, a clustering tree is constructed where each node contains a specific feature (one dimension in the vector) and its corresponding threshold.
The input vector is mapped to the index of a leaf node through simple comparison operations, where each leaf node corresponds to a precomputed centroid (i.e., cluster center) representing the approximate value for data in that region. 
Compared to traditional distance-based clustering methods, this approach can be easily implemented on the constrained dataplane.
Figure~\ref{fig:cache_search} shows an example 
for the input ($x_0 = 3$, $x_1 = 7$). Based on the conditions $x_1 > 5$ and $x_0 \leq 3$, the input is mapped to centroid index (\textit{fuzzy index}) $2$. This index corresponds to the precomputed centroid $(2, 7.5)$. After applying \textsf{Map} $f(X_i) = 0.4X_i + 1$, the approximate results are $(1.8, 4)$.
This approach leverages the continuity of DL  operators (e.g., MatMul and BN), where the operator $f(x)$ remains relatively stable within a small range of input $x$, allowing minor variations in the input without significantly affecting the output~\cite{ziegel2003continuity}.

\begin{figure}[tb]
    \setlength{\belowcaptionskip}{-0.2cm}
	\centering
	\includegraphics[width=0.47\textwidth]{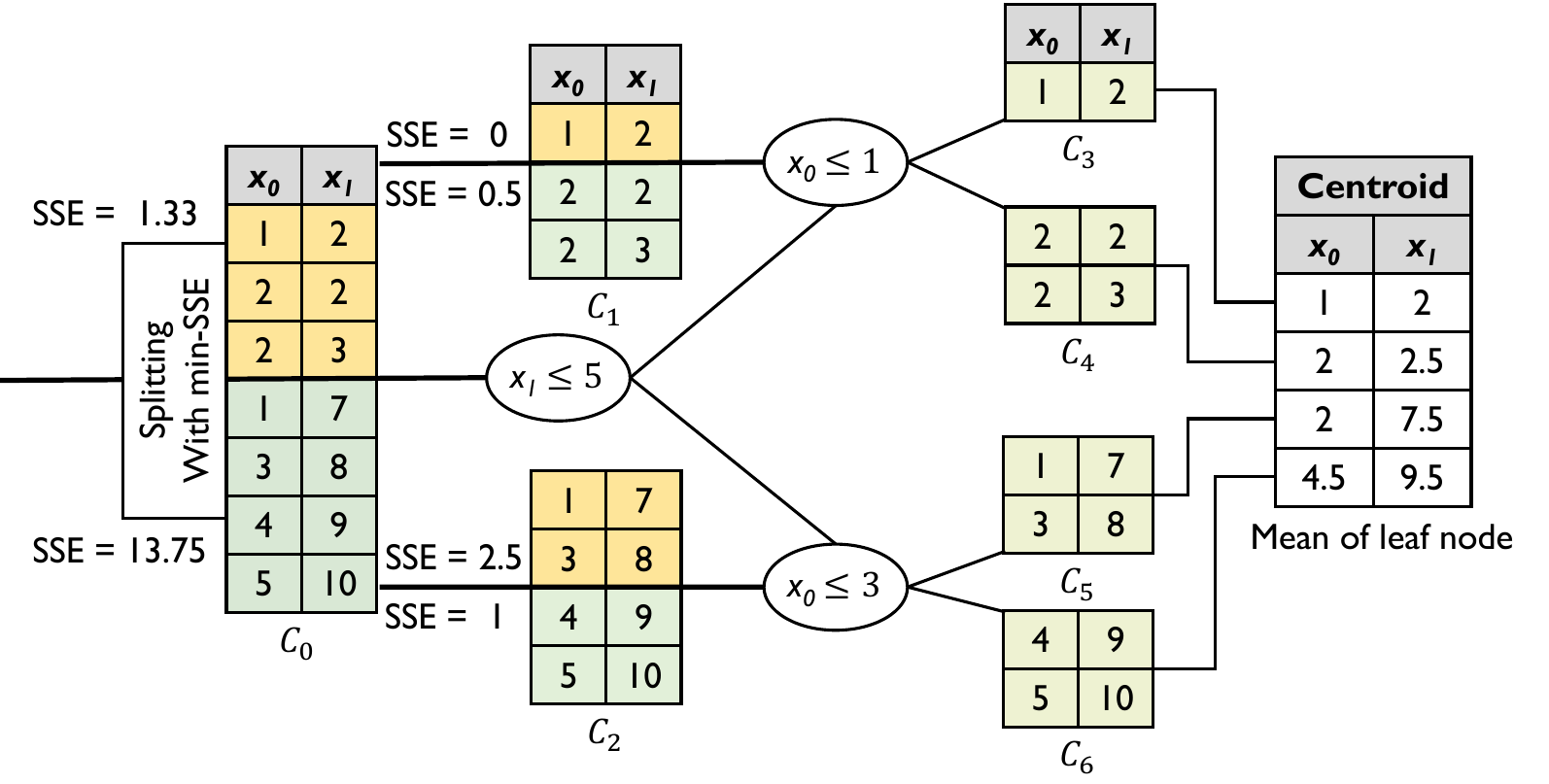}
	\caption{An example of obtaining cluster parameters and centroids from the training dataset in \TIDP.}
	\label{fig:cache_set}
\end{figure}

\noindent\textbf{Parameter Learning.} 
Based on the independent and identically distributed (i.i.d.) assumption of DL \cite{von2011statistical, vapnik1998statistical}, we can learn the parameters (including features and thresholds at the non-leaf nodes, and centroids at the leaf nodes) from the training set for inference.
We adopt a greedy clustering strategy, starting with all training data as a single cluster $C_0$ at the root. 
At each step, we split the current cluster into two sub-clusters by selecting the optimal feature dimension and threshold that minimize the total SSE, maximizing intra-cluster similarity.
For example, as shown in Figure \ref{fig:cache_set}, 
cluster $C_0$ is split along feature $x_1$ at threshold $5$, forming two sub-clusters assigned to the left and right child nodes.
This process continues recursively until the tree reaches the target size. 
Although the greedy strategy does not guarantee a global optimum, it provides a near-optimal split, suitable for efficient dataplane implementation. The centroid of each cluster is computed as the mean vector of its feature dimensions. For instance, the centroid of cluster $C_6$ (4.5, 9.5) is the average of (4, 9) and (5, 10).

\noindent\textbf{Benefits of Fuzzy Matching.}
Compared to storing precomputed input-output mappings for each input unit on the dataplane, fuzzy matching offers four key advantages:
\begin{itemize}[itemsep=0.2em, leftmargin=10pt]
   \item Storage Efficiency: traditional methods suffer from exponential storage growth as the number or bit-width of operands increases. For example, a binary operation (e.g., Hadamard product) or two 8-bit inputs requires $2^{16}$ table entries. Fuzzy matching avoids storing all possible input-output pairs, drastically reducing storage overhead and enhancing scalability.
   \item Lookup Reduction: fuzzy matching enables a single table lookup to cover multiple input units, substantially reducing the number of table lookups and improving memory access bandwidth utilization.
   \item Primitive Fusion: fuzzy matching significantly enhances the capability of \textit{Advanced Primitive Fusion} (see \S\ref{subsec:primitive_fusion}).
   \item Flow Scalability: fuzzy matching supports concurrent flow scalability by storing fuzzy indexes of per-flow features instead of raw data (see \S\ref{subsec:scalability}).
\end{itemize}

\subsection{Primitive Fusion}\label{subsec:primitive_fusion}

In many systems, fusion is employed to optimize resource utilization \cite{fusion_microsecond, fusion_tvm, fusion_zheng}. 
Similarly, we further optimize our primitive implementation through \textit{Primitive Fusion}, which focuses on compressing multiple operations into a single table lookup, thereby improving resource utilization.

\begin{figure}[tb]
	\centering
	\includegraphics[width=0.4\textwidth]{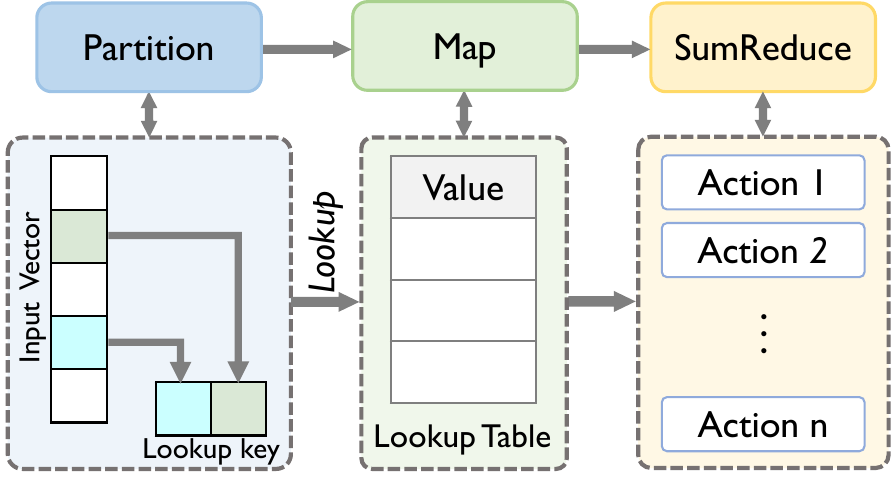}
	\caption{Correspondence between the MAT abstraction and primitives.}
	\label{fig:integration}
    \vspace{-0.2cm}
\end{figure}

As shown in Figure \ref{fig:integration}, an MAT firstly extracts specific fields from the input vector for \textsf{Partition}, and then performs table lookups to retrieve precomputed results of \textsf{Map} primitives, followed by executing corresponding Actions on these results, such as \textsf{SumReduce} primitives.
This process aligns with 
$\textsf{SumReduce}\bigl(\textsf{Map}(\mathcal{F}, \textsf{Partition}(X))\bigr)$,
where $X$ denotes the input vector, and $\mathcal{F}$ represents a set of functions that are applied individually to each partitioned group.

\noindent \textbf{Basic Primitive Fusion.} 
We propose a general approach to fuse primitives \textbf{without modifying the model architecture}. Specifically, we introduce two simple techniques to realize this approach:

(1) Linear Reordering.
If a \textsf{SumReduce} is followed by a \textsf{Map} whose function \(f\) satisfies the linearity property \(f(a + b) = f(a) + f(b)\), we can swap the order of \textsf{SumReduce} and \textsf{Map}. This preserves correctness because applying \(f\) on each partition and then summing is equivalent to summing first and then applying \(f\), provided that \(f\) is linear.

(2) Merging Consecutive \textsf{Map} Primitives.  
Because each \textsf{Map} function applies independently to each partition, consecutive \textsf{Map} operations can be merged into a single \textsf{Map}.

\begin{figure}[tb]
	\centering
	\includegraphics[width=0.47\textwidth]{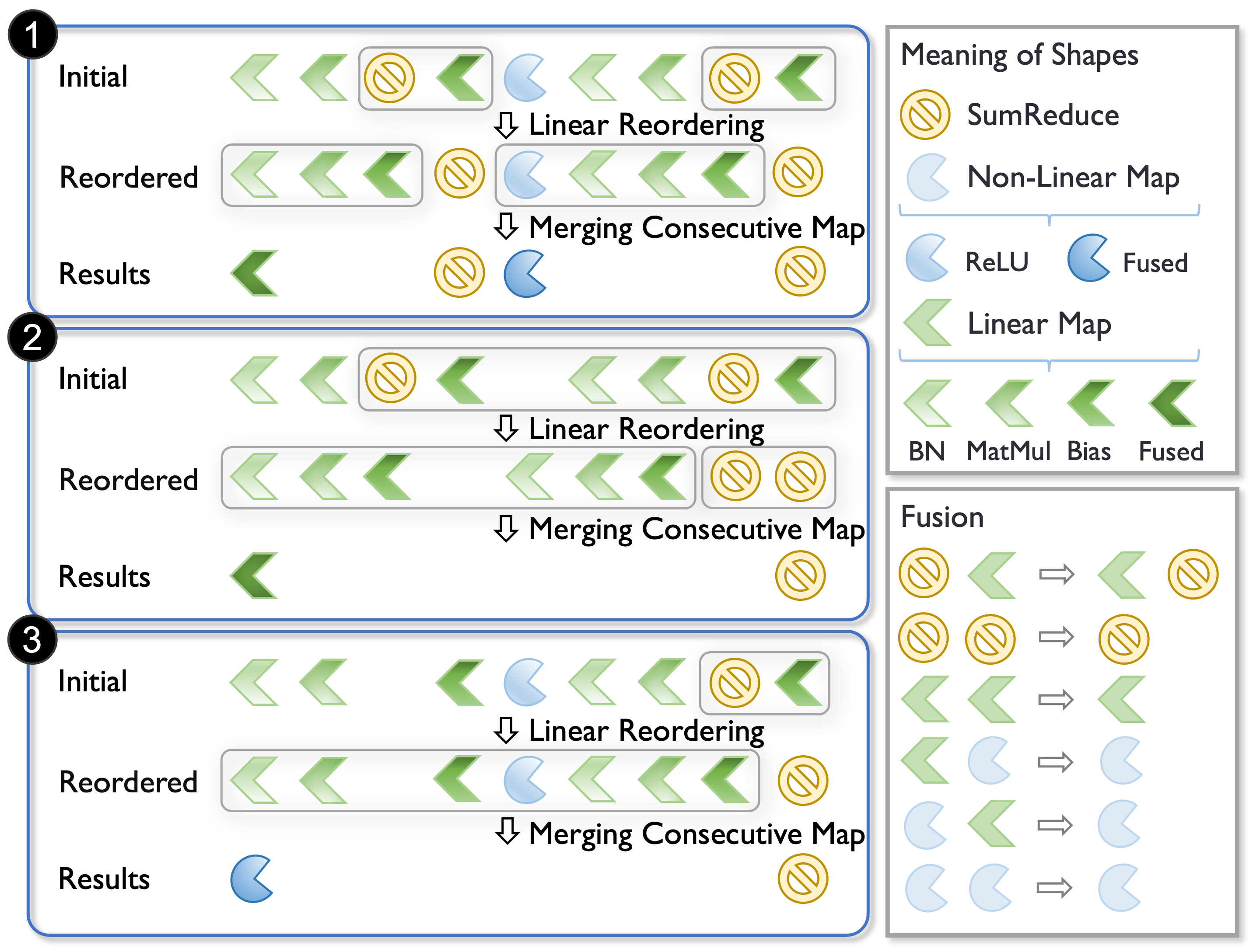}
	\caption{Primitive Fusion techniques: \ding{182} Basic Primitive Fusion; \ding{183} Advanced Primitive Fusion with Removal of Nonlinear Mappings; \ding{184} Advanced Primitive Fusion with Reduction of SumReduce.}
	\label{fig:fusion}
     \vspace{-0.2cm}
\end{figure}

By leveraging these techniques, Pegasus can fuse complex sequences of primitives without altering the underlying model, thereby enabling more efficient inference pipelines. For example, consider an MLP with two hidden layers, where each hidden layer includes:
(1) a BN layer that applies an element-wise linear transform ($\gamma \cdot \frac{x - \mu}{\sigma} + \beta$),
(2) a FC layer that performs MatMul plus bias addition,
(3) a ReLU activation defined as $\max(0, x)$.
The layout of the MLP is shown in the "initial" state of Figure \ref{fig:fusion}~\ding{182}. 
By leveraging basic primitive fusion, we are able to compress seven table lookups into just two (Fused Maps and Fused Non-Linear Maps in the "results" state), thereby eliminating five lookup operations and significantly reducing computational overhead.

\noindent \textbf{Advanced Primitive Fusion.}
To further reduce the overhead associated with table lookups, we propose \textbf{two modifications to the model architecture}. As illustrated in Figure \ref{fig:fusion}, the key to achieving deeper fusion lies in addressing the nonlinear mappings and SumReduce operations. 

(1) Removal of Nonlinear Mappings.
As shown in Figure~\ref{fig:fusion}~\ding{183}, by eliminating all nonlinear mappings from the model, we can compress the entire process into a single table lookup, regardless of the number of intermediate linear mappings. However, while this approach is highly efficient, purely linear models often struggle to capture complex patterns and relationships within the data, potentially leading to a significant drop in accuracy.

(2) Reduction of SumReduce Operations.
As shown in Figure~\ref{fig:fusion}~\ding{184}, by retaining only the final SumReduce operation and removing all others, we can also condense the model into a single table lookup. This method, similar to Neural Additive Models (NAM)~\cite{agarwal2021nam}, can effectively capture complex and nonlinear relationships within each segment. 
This method benefits from \textit{fuzzy matching}, which allows for more data within each segment.
The outputs from each partition are then aggregated through SumReduce, allowing for a straightforward yet comprehensive integration of global information while maintaining the independence of individual sub-models.

\subsection{Mapping Optimization}\label{subsec:optimization}
Primitive Fusion allows us to cluster inputs only before the fused large operators, replacing the original inputs with centroids to reduce the pressure on the dataplane. However, this approach \textbf{inevitably introduces approximation errors}. 
To ensure the mapping table more accurately aligns with the model’s actual output, we employ \textbf{backpropagation} to \textbf{dynamically adjust} the stored centroids and cluster parameters, making it closer to the ideal performance.

\noindent \textbf{Backpropagation.}
\TIDP first trains an initial model on the training dataset to generate cluster parameters and centroids in the mapping table.
Subsequently, \TIDP constructs mapping tables and performs centroid assignment within the model using the technique from Zhang \cite{zhang2021yet}. 
This method allows us to simulate the centroid assignment process in the model through matrix operations.
Backpropagation is then applied to fine-tune the cluster parameters and centroids, improving their alignment with the model’s output, thereby reducing errors and minimizing the impact on overall performance.

\noindent \textbf{Adaptive Fixed-Point Quantization.}
During the inference process, the fixed-point positions of inputs and outputs can differ, especially when there are significant differences in numerical ranges (e.g., input range [-100, 100] versus output range [0, 5]). Some inference hardware employs Post-Training Static Quantization \cite{pytorch_tensorrt_ptq}, which pre-defines the fixed-point positions for each layer’s weights and activations based on known numerical ranges. This method helps maximize register bit-width utilization and improve numerical precision during inference.

In \TIDP, since the mapping table stores operations at full precision, we only need to perform fixed-point quantization on the final outputs before the \textsf{SumReduce} primitive.
We pre-calculate the fixed-point positions and store the corresponding outputs in a mapping table.
This approach allows \textsf{Map} primitives to handle inputs and outputs with different fixed-point positions, enhancing precision, particularly when there is a mismatch in numerical ranges. By optimizing in this manner, \TIDP flexibly processes data across varying ranges without sacrificing computational accuracy.
\section{Deep Learning Operators}\label{DLO}

\begin{table}[tb]
\centering
\footnotesize
\begin{tabular}{l|l}
\hline
\rowcolor[HTML]{EFEFEF} 
{\color[HTML]{333333} DL Layers} & {\color[HTML]{333333} DL Operators}                                                                                                                                                                                       \\ \hline
Emb                              & Embedding Lookup                                                                                                                                                                                                          \\ \hline
FC                               & \begin{tabular}[c]{@{}l@{}}Matrix Multiplication (Weighted Aggregation)\\ Bias Addition (Element-wise Transformation)\end{tabular}                                                                                        \\ \hline
Conv                             & Convolution (Weighted Aggregation)                                                                                                                                                                                        \\ \hline
Act                              & \begin{tabular}[c]{@{}l@{}}ReLU, tanh (Element-wise Transformation)\\ Softmax (Multi-Input Operation)\end{tabular}                                                                                                        \\ \hline
Norm                             & \begin{tabular}[c]{@{}l@{}}Batch Normalization (Element-wise Transformation)\\ Layer Normalization (Multi-Input Operation)\end{tabular}                                                                                   \\ \hline
Pool                             & Pooling (Multi-Input Operation)                                                                                                                                                                                           \\ \hline
Rec                              & \begin{tabular}[c]{@{}l@{}}Matrix Multiplication (Weighted Aggregation) \\ Bias Addition (Element-wise Transformation)\\ tanh, Sigmoid (Element-wise Transformation)\\ Hadmard (Element-wise Transformation)\end{tabular} \\ \hline
\end{tabular}
\vspace{0.5mm}
\caption{Operators in DL layers.
}
\label{tab:primary_operations}
     \vspace{-0.3cm}
\end{table}

In deep learning (DL), layers are the building blocks of neural networks, each designed to perform specific transformations on the input. DL layers are typically constructed from a set of DL operators, as outlined in Table \ref{tab:primary_operations}, which maps layers to their corresponding operators. In this section,  we explain how \TIDP primitives can be used to implement these DL operations. All references to DL layers are focused on the inference phase.

\noindent \textbf{\textbullet\ Embedding Lookup.}
Embedding Lookup is commonly used in embedding layers during inference, mapping discrete input indices to dense vectors. It can be viewed as an indexing function \( f(x) = E[x] \), efficiently implemented using the \textsf{Map} primitive.

\noindent \textbf{\textbullet\ Element-wise Transformation.}
Element-wise Transformation refers to operations performed independently on each element of the input, making it naturally suitable for implementation using the \textsf{Map} primitive. 
During inference, most parameters, such as weights, biases, and other model parameters, are known in advance. These can be treated as constants, part of the function rather than inputs, reducing the computational overhead during the mapping process.

\noindent \textbf{\textbullet\ Weighted Aggregation.}
Weighted Aggregation is the most computationally intensive operation in DL~\cite{dean2018new_weighted, sze2017efficient_weighted, goodfellow2016deep_weighted}, generating output by performing element-wise multiplication between input elements and their corresponding weights, followed by summing the results.
This operation can be Partitioned into multiple parts, with each part processed using the \textsf{Map} primitive and the corresponding weights. The result vector can be retrieved directly through a single table lookup, and the final output is obtained by applying the \textsf{SumReduce} primitive to aggregate the results.

\noindent \textbf{\textbullet\ Multi-Input Operation.}
Multi-Input Operation refers to computations where an element’s output depends on multiple input elements. 
These inputs may be too numerous to fit into a single partition due to combinatorial explosion.
There are two common ways to implement this operation. 
The first method uses the \textsf{Map} primitive to process each partition, then apply the \textsf{SumReduce} primitive to aggregate their influence on the output, followed by \textsf{Map} primitives to operate on the aggregated result and produce the final output.
For example, Softmax (defined as $\text{Softmax}(x_i) = \frac{e^{x_i}}{\sum_j e^{x_j}}$) involves a \textsf{Map} primitive to exponentiate each element  $e^{x_i}$, followed by a \textsf{SumReduce} primitive to sum these values  $\sum e^{x_i}$, and a final \textsf{Map} primitive to normalize each element by this sum  $\frac{e^{x_i}}{\sum_j e^{x_j}}$. 
The second method uses consecutive \textsf{Map} primitives to progressively compute operations between multiple elements. For instance, average pooling requires several \textsf{Map} primitives to iteratively compute the average value, yielding the final result.

\section{Implementation}\label{MRD}

\TIDP is generalizable to commodity programmable swi-tches, such as PISA-based~\cite{p4} and Trio-based~\cite{yang2022trio} switches, which support the P4 language. 
This generalizability stems from \TIDP’s reliance solely on comparisons, table lookups, and additions. To demonstrate its practicality, we have implemented \TIDP on the PISA switch.

\subsection{Fuzzy Matching Implementation} 
In \TIDP, the input traverses the clustering tree to obtain the \textit{fuzzy index}. This process requires a multi-level comparator, which is not natively supported by PISA-based switches.
To address this, we use the numerical range of values to represent the leaf nodes of the clustering tree. This approach leverages range matching to facilitate the implementation of the mapping table.
To efficiently convert these ranges into ternary rules, we introduce the Consecutive Range Coding (CRC) algorithm \cite{netbeacon}, which enables the effective transformation of numerical ranges into ternary rules. 

\subsection{\TIDP Syntax}

\begin{figure}[tb]
    \setlength{\abovecaptionskip}{0.3cm}
    \setlength{\belowcaptionskip}{-0.2cm}
	\centering
	\includegraphics[width=0.47\textwidth]{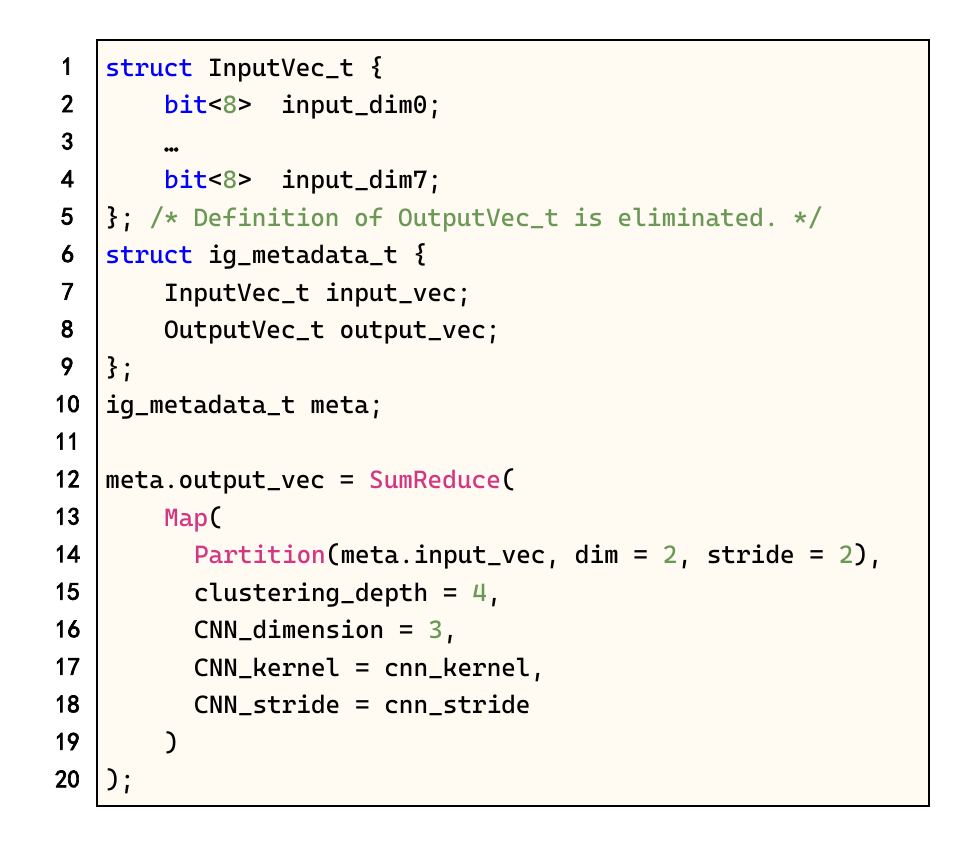}
	\caption{\TIDP Syntax.}
	\label{fig:syntax}
\end{figure}

To facilitate the implementation of various DL models on the \TIDP framework, we have designed a specialized syntax called \TIDP Syntax. Figure~\ref{fig:syntax} illustrates the proposed syntax, which provides a high-level abstraction for defining and configuring DL models.
To support the translation of \TIDP Syntax into P4 language, we developed a translation tool. This tool significantly reduces programming complexity, allowing developers to focus on high-level logic design without delving into the intricacies of low-level P4 code. 

Specifically, our \textsf{Pegasus} Syntax maintains a consistent form with the primitives. 
In \textsf{Partition} phase, input data and its partitioning rules are explicitly specified. The partitioned data in each segment is used to perform \textsf{Map} operations.
In the \textsf{Map} phase, we define the depth of the clustering tree and a series of CNN parameters to determine the output dimensions for each group of inputs. 
The translator automatically calculates the output dimensions based on these parameters. 
This design is motivated by the fact that certain operations, such as the convolution process in CNNs, are partially connected. Reducing the output dimensions of the \textsf{Map} primitives can effectively minimize table resource overhead.
The specific allocation of hardware resources is automatically handled by the translator. 

\subsection{Implemented Neural Networks}
\label{subsec:implemented_NN}

We implemented the following six representative DL models\footnote{In addition to the six DL models mentioned in the text, their variants can also be implemented using these methods. Note that large models, such as Transformers, cannot be supported due to resource limitations.}
within \TIDP, all of which utilize the fuzzy matching and Basic Primitive Fusion. Additionally, we applied the Advanced Primitive Fusion technique in CNN-M, CNN-L and AutoEncoder to enable larger model scales with lower overhead, achieving improved accuracy.

\noindent \textbf{MLP-B.}
MLP is well-suited for handling high-dimensional data, making it particularly effective in processing statistical features. 
We implemented a basic MLP model (MLP-B) that operates on statical features, including flow-level and packet-level features. 
However, It’s hard to extract effective statistical features for MLP in the dataplane.
For example, calculating averages is challenging on programmable switches, while using cumulative sums can lead to overfitting to large flows. 
To ensure fairness, we only use the maximum and minimum packet lengths and inter-packet delays (IPD) as flow-level information.
Our MLP-B consists of three hidden layers, each comprising a sequence of Batch Normalization, a FC layer, and a ReLU activation function.

\noindent \textbf{RNN-B.}
RNNs are well-suited for capturing temporal dependencies in sequential data, making them particularly effective for handling time-series features. Our implementation is based on the windowed binary RNN design in BoS~\cite{bos}, which processes multiple time steps on the switch to capture sequential dependencies without requiring hidden state write-backs. It classifies packets based on the sequence of packet lengths and inter-packet delays (IPD).
The RNN-B model consists of an Emb layer, a tanh activation function, and multiple FC layers.

\noindent \textbf{CNN-B, CNN-M, and CNN-L.}
The one-dimensional CNN demonstrates unique advantages in processing windowed sequence data \cite{jacovi2018cnn_text}. We implemented three CNN models: CNN-B (basic), CNN-M (medium), and CNN-L (large), with increasing model complexity and scalability.
CNN-B serves as the baseline model, employing only the Basic Primitive Fusion technique. It uses packet length and IPD sequences as input features.
CNN-S extends CNN-B by incorporating Advanced Primitive Fusion, enabling larger model scales with lower overhead.
CNN-L builds on CNN-S by further leveraging Advanced Primitive Fusion to support even larger model sizes and input scales. This enables CNN-L to extract 60 raw bytes from each packet as a raw packet sequence.
All three models are based on the textcnn architecture proposed by Zhang et al. \cite{zhang2015cnn_text}, consisting of multiple Conv layers, FC layers, Pool layers, and ReLU activation functions.

\noindent \textbf{AutoEncoder.} 
Autoencoders are effective for unsupervised anomaly detection by learning compact representations and reconstructing input data. Our implementation uses mean absolute error (MAE) to calculate reconstruction error, which is then used to determine whether a flow is anomalous.
The Autoencoder model consists of an Emb layer and multiple FC layers for encoding and decoding. Each FC layer is preceded by a Batch Normalization layer and followed by a ReLU activation function.

\section{Evaluation}

Our evaluation addresses the following questions:
(i) Whether \TIDP can achieve higher accuracy and generality in supporting a variety of DL models? (\S \ref{subsec:performance}).
(ii) How scalable is \TIDP across model size, input scale, and the number of simultaneous flows? (\S \ref{subsec:scalability}).
(iii) Whether the unsupervised models implemented by \TIDP can effectively defend against real-world unknown attack traffic? (\S \ref{subsec:unsupervised}).
(iv) What advantages DL model implementations on the dataplane offer compared to those on the control plane? (\S \ref{subsec:compare_control}).

\subsection{Experiment Setup}\label{subsec:experiment_setup}

\noindent \textbf{Testbed Setup.}
We implemented \TIDP using P4~\cite{p4} on a Barefoot Tofino 2 programmable switch~\cite{tofino2}, connected to two Linux servers. One server replays pcap files via tcpreplay, while the other server receives packets from the programmable switch.

\noindent \textbf{Traffic Classification Datasets.}
We use three publicly available and widely used traffic classification datasets, which are also utilized in BoS~\cite{bos}: 
(i) PeerRush \cite{rahbarinia2013peerrush}: This dataset contains traffic generated by P2P applications, categorized into three classes (eMule, uTorrent, and Vuze).
(ii) CICIOT2022 (CICIOT) \cite{ciciot}: This dataset contains traffic collected from IoT devices in different working states, categorized into three classes: Power, Idle, and Interact.
(iii) ISCXVPN2016 (ISCXVPN) \cite{vpn}: This dataset consists of VPN-encrypted network traffic, categorized into seven classes (Email, Chat, Streaming, FTP, VoIP, P2P). 
For each dataset, we selected 75\% of the flows (identified by five-tuple) from each class to train the DL models, 10\% for validation, and 15\% for testing.

\noindent \textbf{Baselines.}
We implemented N3IC~\cite{n3ic}, Leo~\cite{leo}, and BoS~\cite{bos}, using the largest model configurations specified in their respective papers. Among these, Leo and BoS were deployed on the switch, while N3IC was evaluated through software simulation because the largest models in their papers could not be implemented on the switch.
It is important to note that our evaluation focuses solely on the accuracy of the models themselves. We did not employ common optimization techniques, such as those in BoS, which enhance accuracy by aggregating predictions from multiple packets within a flow and offloading hard-to-classify cases to the control plane via the Integrated Model Inference System (IMIS).

\noindent \textbf{Metrics.}
Consistent with prior works \cite{bos, netbeacon}, we use packet-level macro-accuracy, defined as the average F1-score across different classes, to evaluate model accuracy. Unless otherwise specified, all accuracy measurements in the evaluation refer to macro-accuracy. Additionally, we report the overall Precision (PR) and Recall (RC) to provide a comprehensive evaluation of the models.

\renewcommand{\arraystretch}{1.10}
\begin{table*}[tb]
\centering
\footnotesize
\resizebox{\textwidth}{!}{
\begin{tabular}{c|c|c|ccc|ccc|ccc}
\hline
\multirow{2}{*}{Method} &
  \multirow{2}{*}{\begin{tabular}[c]{@{}c@{}}Input\\ Scale (b)\end{tabular}} &
  \multirow{2}{*}{\begin{tabular}[c]{@{}c@{}}Model\\ Size (Kb)\end{tabular}} &
  \multicolumn{3}{c|}{PeerRush} &
  \multicolumn{3}{c|}{CICIOT} &
  \multicolumn{3}{c}{ISCXVPN} \\ \cline{4-12} 
                               &     &      & PR     & RC     & F1     & PR     & RC     & F1     & PR     & RC     & F1     \\ \hline
Leo \cite{leo} (Decision Tree) & 128 & -    & 0.8720 & 0.8776 & 0.8728 & 0.7910 & 0.8072 & 0.7848 & 0.7338 & 0.7797 & 0.7475 \\
N3IC \cite{n3ic} (binary MLP)  & 128 & 24.4 & 0.8217 & 0.8308 & 0.8241 & 0.7855 & 0.7877 & 0.7745 & 0.6688 & 0.6521 & 0.6388 \\
MLP-B                          & 128 & 34.3 & 0.8823 & 0.8826 & 0.8823 & 0.8555 & 0.8615 & 0.8581 & 0.7676 & 0.7552 & 0.7574 \\ \hline
BoS \cite{bos} (binary RNN)    & 18  & 25.6 & 0.8677 & 0.8696 & 0.8678 & 0.8311 & 0.8253 & 0.8276 & 0.7033 & 0.7089 & 0.6907 \\
RNN-B                          & 128 & 10.9 & 0.9083 & 0.9100 & 0.9090 & 0.8707 & 0.8708 & 0.8707 & 0.7848 & 0.7658 & 0.7617 \\
CNN-B                          & 128 & 11.4 & 0.9051 & 0.9069 & 0.9057 & 0.8861 & 0.8657 & 0.8659 & 0.7706 & 0.7600 & 0.7520 \\ \hline
CNN-M                          & 128 & 974  & 0.9201 & 0.9220 & 0.9207 & 0.8821 & 0.8839 & 0.8829 & 0.7942 & 0.7897 & 0.7780 \\ \hline
\textbf{CNN-L} &
  \textbf{3840} &
  \textbf{6083} &
  \textbf{0.9967} &
  \textbf{0.9966} &
  \textbf{0.9966} &
  \textbf{0.9391} &
  \textbf{0.9377} &
  \textbf{0.9380} &
  \textbf{0.9868} &
  \textbf{0.9877} &
  \textbf{0.9872} \\ \hline
\end{tabular}
}
\vspace{0.5mm}
\caption{Comparison of classification accuracy across different methods.}
\vspace{-0.3cm}
\label{tab:performance}
\end{table*}

\begin{figure*}[tb]
    \centering
    \begin{minipage}[b]{0.23\textwidth} 
        \centering
        \includegraphics[width=1\textwidth]{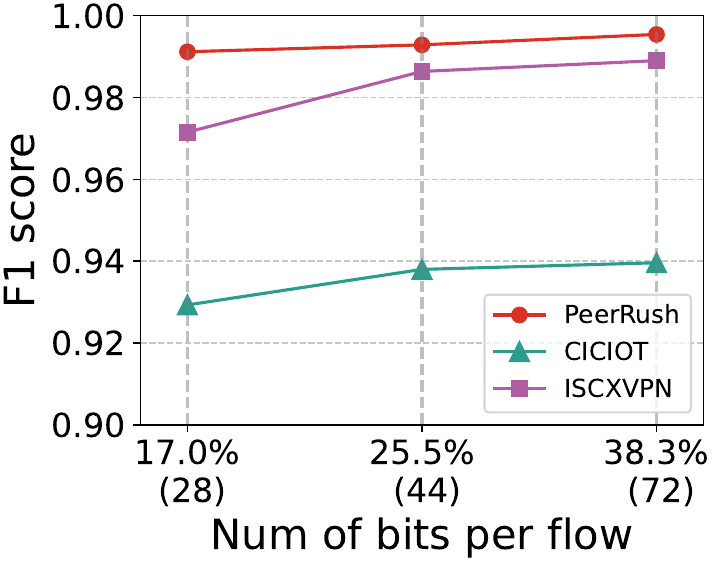}\label{fig:perflow_impact_f1}
        \caption{Impact of per-flow storage usage on classification accuracy.}
        \label{fig:perflow_impact}
    \end{minipage}
    \hfill
    \begin{minipage}[b]{0.74\textwidth} 
        \centering
        \subfigure[PeerRush]{\includegraphics[width=0.31\textwidth]{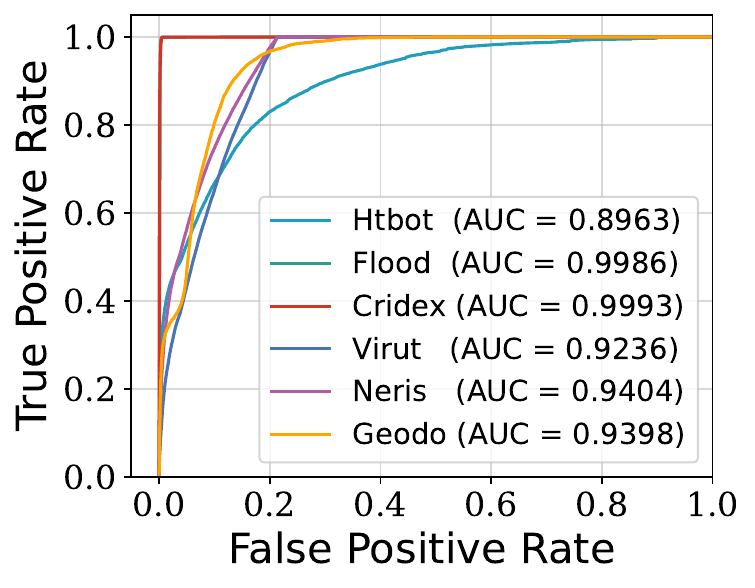}\label{fig:unsupervised_peerrush}}
        \subfigure[CICIOT]{\includegraphics[width=0.31\textwidth]{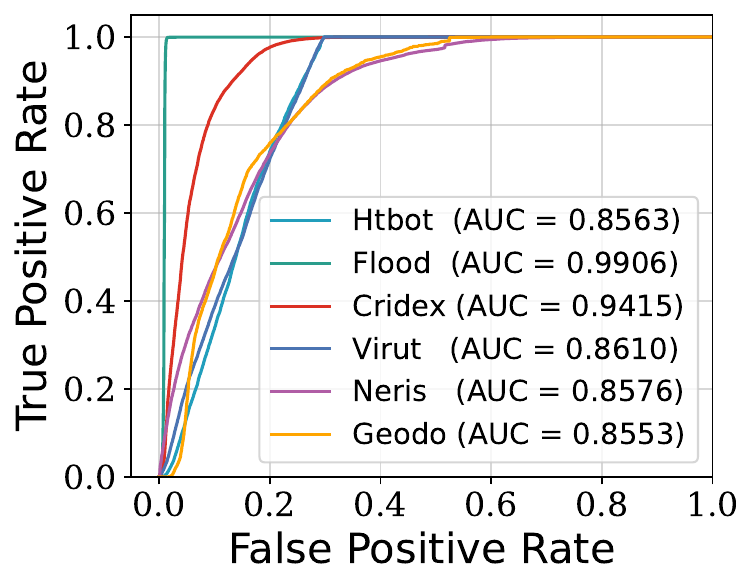}\label{fig:unsupervised_ciciot}}
        \subfigure[ISCXVPN]{\includegraphics[width=0.31\textwidth]{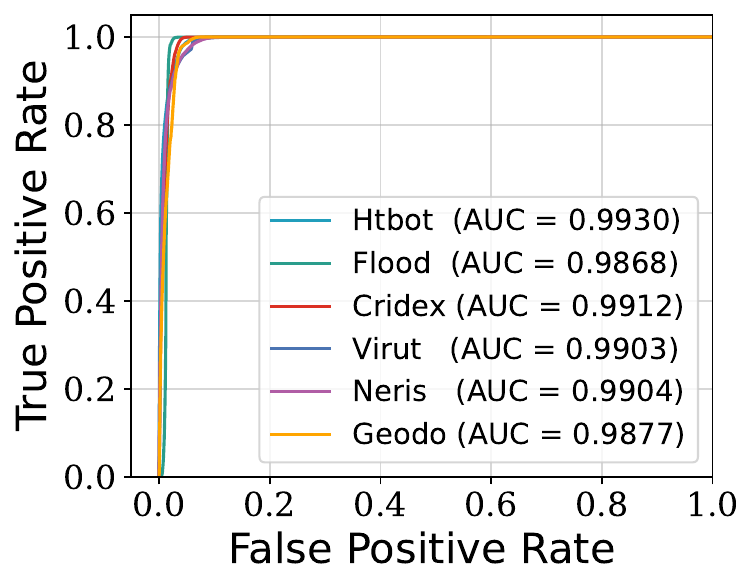}\label{fig:unsupervised_vpn}}
        \caption{ROC curves across different datasets.}
        \label{fig:unsupervise}
    \end{minipage}
\end{figure*}
\begin{table}[tb]
\small
\centering
\begin{tabular}{c|cccc}
\hline
Models      & \begin{tabular}[c]{@{}c@{}}Stateful\\ bits/flow\end{tabular} & SRAM   & TCAM    & Bus     \\ \hline
Leo         & 80                                                           & 2.44\% & 21.67\% & 3.55\%  \\
BoS         & 72                                                           & 2.81\% & 0\%     & 0.74\%  \\ \hline
MLP-B       & 80                                                           & 7.75\% & 12.92\% & 29.45\% \\
RNN-B       & 240                                                          & 7.38\% & 23.33\% & 33.36\% \\
CNN-B       & 72                                                           & 5.56\% & 7.08\%  & 13.16\% \\
CNN-M       & 72                                                           & 3.50\% & 6.67\%  & 3.98\%  \\
CNN-L       & 44                                                           & 7.12\% & 13.33\% & 7.11\%  \\ \hline
AutoEncoder & 240                                                          & 5.06\% & 7.92\%  & 7.23\%  \\ \hline
\end{tabular}
\vspace{2mm}
\caption{Hardware resource utilization for different methods.}
\label{tab:resource}
\vspace{-0.3cm}
\end{table}

\subsection{Accuracy Comparison and Analysis}
\label{subsec:performance}

In this section, we compare the accuracy of Leo, N3IC, and MLP-B used the same statical features; BoS, RNN-B and CNN-B using the same raw packet sequence. 
Detailed analysis of CNN-S and CNN-L is deferred to \S\ref{subsec:scalability}.
We summarize all classification accuracy results in Table \ref{tab:performance}.

\noindent \textbf{Statical Features.}
As shown in Table \ref{tab:performance}, MLP-B achieves better accuracy than N3IC, with improvements ranging from 5.8\% to 11.9\%, despite the two models having similar sizes. This illustrates the accuracy degradation caused by the full-model binarization in N3IC, particularly in the absence of Norm and Act layers. 
In contrast, \TIDP uses full-precision model weights and fixed-point activations to enhance accuracy. This design choices allow \TIDP to maintain its accuracy advantage even under the fuzzy matching-induced errors.

Compared to Leo, \TIDP achieved a 7.3\% accuracy improvement on the CICIOT dataset. This advantage may stem from the complex relationships among CICIOT features, which an MLP model of this scale can capture more effectively than tree-based approaches.
However, the improvement remains modest on the PeerRush and ISCXVPN tasks, with only an average 1.0\% gain. 
This is consistent with the fact that decision trees perform well on statistical features. Nevertheless, DL models excel at processing raw packet sequences, a capability we will demonstrate through the CNN-L implementation~(see~\ref{subsec:scalability}).

\noindent \textbf{Raw Packet Sequence.}
As shown in Table \ref{tab:performance}, despite using full-precision model weights, the BoS still exhibits a 4.1\% to 7.1\% lower accuracy than RNN-B across the three traffic classification tasks. This confirms the impact of input and output binarization on model accuracy.

Additionally, the CNN-B model demonstrates comparable accuracy to the RNN-B model but performs slightly lower, with an average gap of 0.6\%. This may be attributed to RNN’s superior ability to capture sequential dependencies under the same model size.

\subsection{Scalability Evaluation} \label{subsec:scalability}
MLP models are constrained by the switch’s limited ability to extract complex statistical features, while RNN models face implementation challenges on the switch due to the requirement for sequential execution over multiple time steps. 
Given these limitations, we select CNN models to evaluate the impact of scalability on classification accuracy, as their accuracy is significantly affected by model size.

\noindent \textbf{Model Scale Scalability.}
As shown in Table \ref{tab:performance}, as the model size increases, CNN-M achieves accuracy improvements of 1.5\% to 2.6\% over CNN-B, while outperforming RNN-B by 1.2\% to 1.6\%.
This improvement is not proportional to model size, as larger models face diminishing returns due to feature saturation and dataset complexity limits.
Nevertheless, CNN-M achieves these gains with lower overhead compared to CNN-B (see \S\ref{subsec:unsupervised}). By leveraging Advanced Primitive Fusion, CNN-M significantly reduces the number of tables, optimizing resource utilization.

To further improve traffic classification accuracy, we expanded the feature set and model size. With this enhancement, CNN-L demonstrates exceptional accuracy, achieving 99.66\%, 93.80\%, and 98.72\% on the PeerRush, CICIOT, and ISCXVPN datasets, respectively. This represents an improvement of 7.2\% to 23.5\% over CNN-B, and average accuracy gains of 17.2\%, 22.8\% and 17.9\% over Leo, N3IC, and BoS, respectively.

Additionally, CNN-L has a model size of 6083Kb, which is 248x and 237x larger than those of N3IC (24.4Kb) and BoS (25.6Kb), respectively (see \S\ref{subsec:unsupervised}).
CNN-L also supports input sizes of 3840 bits, representing a 29x increase over N3IC (128 bits) and a 212x increase over BoS (18 bits).
Traditional methods struggle to support such large input sizes for two main reasons:
(a)~PISA switches only have 4096-bit Packet Header Vector (PHV), making it difficult to handle such large-scale features while supporting basic functionalities. 
CNN-L benefits from the design of primitives.
Specifically, CNN-L uses \textsf{Partition} to divide the input, distributing the inference process across each packet within the window. Each packet only processes 480 bits of features, allowing CNN-L to successfully implement.
(b)~Excessive per-flow register usage can impact the number of concurrent flows that can be supported (see below).
This demonstrates Pegasus’s exceptional scalability.

\noindent \textbf{Number of Concurrent Flows Supported.}
Storing per-flow features on the switch requires the use of its stateful SRAM resources. Supporting more per-flow features reduces the number of concurrent flows that can be managed, which is why previous works did not support features at the same scale as CNN-L on the dataplane.

However, \TIDP achieves this at a low cost. For scenarios requiring a larger number of per-flow features, such as CNN-L, \TIDP first uses a neural network to extract high-level, refined features from each packet, reducing the per-flow storage needed. These features, similar to those in CNN-S and CNN-B that require less per-flow storage, can be further compressed through fuzzy matching, which maps the features into fuzzy indexes in the mapping table, significantly reducing storage overhead.

Figure \ref{fig:perflow_impact} summarizes how classification accuracy varies with the per-flow storage overhead, where the X-axis corresponds to the required SRAM overhead to support 1 M flows for different per-flow storage sizes. 
The CNN-L model uses 48 bits per flow, including 16 bits for the previous packet timestamp (used for IPD calculation) and 4 bits for the fuzzy index extracted from each packet (for a window size of 8, the features of 7 packets need to be stored). Additionally, a 28-bit version of CNN-L removes the IPD feature, while a 72-bit version extracts 8 bits for the fuzzy index from each packet \footnote{In fact, since PISA switches do not support 4-bit registers, we actually used 4 8-bit registers to replace the 7 4-bit registers.}.
Even with 28 bits of per-flow storage, the model achieves classification accuracies of 99.1\%, 92.9\%, and 97.2\% on the PeerRush, CICIOT, and ISCXVPN datasets, respectively. Compared to Leo, N3IC, and BoS, it improves the average accuracy by 16.2\%, 21.8\%, and 16.9\%, respectively. Moreover, the 28-bit per-flow storage usage is significantly lower than BoS's 72-bit usage and the 80-bit usage of Leo and N3IC (see \S\ref{subsec:unsupervised}).

\subsection{Unsupervised Malicious Traffic Detection Evaluation} \label{subsec:unsupervised}

Previous works have predominantly focused on leveraging learning models to classify traffic on the dataplane under scenarios with abundant labeled data. However, in real-world networks, attacks often come from unknown traffic, such as zero-day attacks. It is unrealistic to anticipate such attack traffic in advance and train supervised models accordingly.
Detecting unknown traffic through unsupervised models is challenging as it requires extracting multiple complex features from the traffic, reconstructing the original inputs using large model structures, and determining whether the traffic is malicious based on reconstruction errors~\cite{mirsky2018kitsune, auc_autoencoder_tang2020zerowall}. This complexity has prevented prior works from addressing this area on the dataplane.

In this section, we validate that the AutoEncoder implemented by \TIDP can utilize a large model structure to extract features from raw packet sequences (packet length and IPD) and identify unknown attack traffic by calculating reconstruction errors using MAE. Specifically, the model leverages knowledge learned in the Emb layer during traffic classification tasks to capture relationships and features from raw packet sequences. These features are reconstructed through the encoder and decoder.

\noindent \textbf{Datasets.}
We use the AutoEncoder to reconstruct traffic on the training sets of the PeerRush, CICIOT, and ISCXVPN datasets. To evaluate the model’s ability to detect unknown attacks, we inject two representative malicious traffic at a 1:4 mixture of attack-to-benign traffic into the testing set: (a)~Malware Attack, including Cridex, Geodo, Htbot, Neris, and Virut, sourced from USTC-TFC2016~\cite{wang2017malware}. (b)~DoS Attack, utilizing SSDP Reflection Flood traffic, collected from Kitsune~\cite{mirsky2018kitsune}.

\noindent \textbf{Metrics.}
We use AUC (AURoC, Area Under the Receiver Operating Characteristic Curve) as metrics, as these are commonly used in existing studies~\cite{auc_deeplog,auc_fu2024detecting,auc_holland2021new,auc_zhu2020you}. AUC measures the model’s ability to distinguish between normal and malicious traffic.

\noindent \textbf{Results.}
As shown in Figure \ref{fig:unsupervise}, the AutoEncoder achieves average AUCs of 95.0\%, 89.4\%, and 99.0\% on the PeerRush, CICIOT, and ISCXVPN datasets, respectively, across different types of malicious traffic. This demonstrates that the AutoEncoder can effectively distinguish normal traffic from anomalous traffic when only normal traffic is available during training.
In practical deployments, programmable switches can dynamically adjust response strategies based on the MAE value and its abnormal fluctuations. For instance, they can enforce traffic rate limits or send real-time alerts to administrators, enabling the system to handle potential malicious traffic or attacks more efficiently.

\noindent \textbf{Hardware Resource Utilization.} \label{hardware_resource}
Unlike the accuracy evaluation, we implemented moderately sized versions of BoS (with a hidden size of 8) and Leo (with 1024 nodes) to assess resource overhead, as models like BoS are inherently designed for small-scale scenarios.

We report the stateful per-flow bit usage, stateless SRAM and TCAM overhead, and Action Data Bus utilization for implementing different methods on the switch in Table \ref{tab:resource}.
In \TIDP, TCAMs are used to retrieve the fuzzy index, the SRAMs are used to store mapping tables, and the Action Data Bus are used to transfer data fetched from SRAM/TCAM.

Compared to CNN-B, CNN-M has a larger model size but lower resource overhead. This is primarily due to the fusion of all intermediate-layer operations through Advanced Primitive Fusion, which improves resource utilization. This effect is even more pronounced in CNN-L. Despite having a model size of 6083Kb, CNN-L only occupies 7.12\% of SRAM and 13.33\% of TCAM. The majority of the model parameters are fused, so they do not occupy storage resources during inference.

Compared to BoS and Leo, CNN-L has a larger resource overhead, which is understandable given its larger model scale and higher numerical precision. However, this drawback is alleviated as the number of concurrent flows increases, because CNN-L has a lower per-flow register usage. As shown in Figure \ref{fig:perflow_impact}, when supporting 1M concurrent flows, the 28-bit per-flow storage version of CNN-L saves 21.3\% of SRAM overhead compared to the 72-bit requirement, thereby significantly mitigating this drawback.

\subsection{Compare With Control Plane DL} \label{subsec:compare_control}

\begin{figure*}[tb]
    \centering
    \subfigure[PeerRush]{\includegraphics[width=0.24\textwidth]{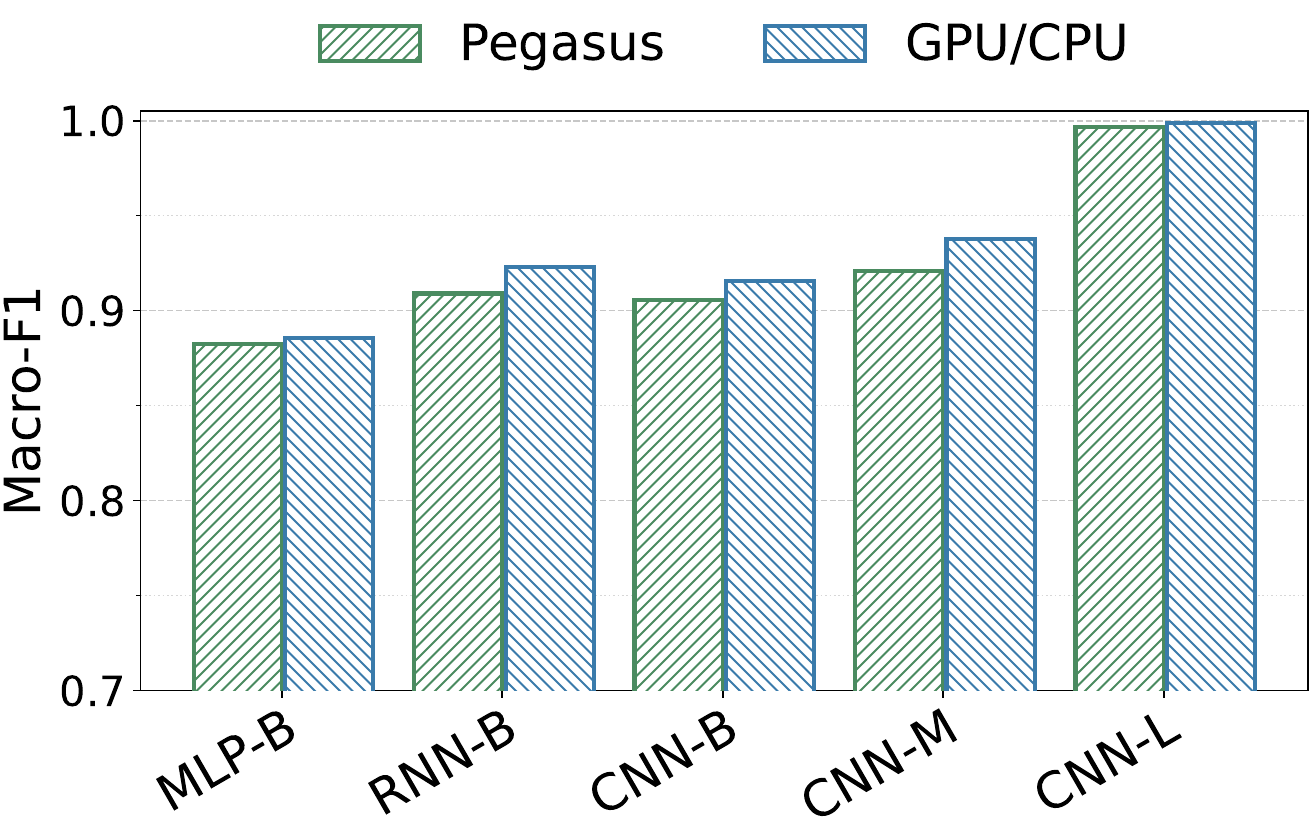}\label{fig:degration_PeerRush}}
    \subfigure[CICIOT]{\includegraphics[width=0.24\textwidth]{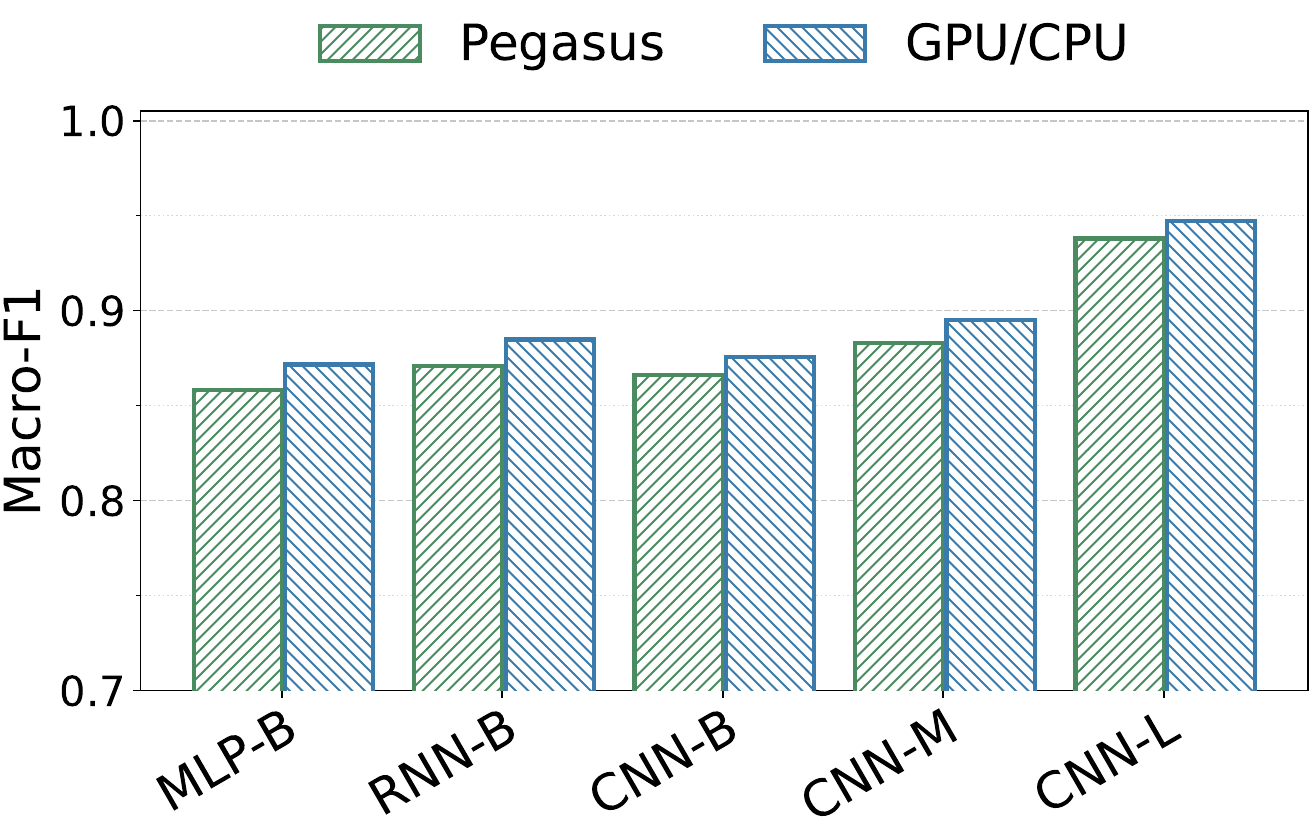}\label{fig:degration_CICIOT}}
    \subfigure[VPN]{\includegraphics[width=0.24\textwidth]{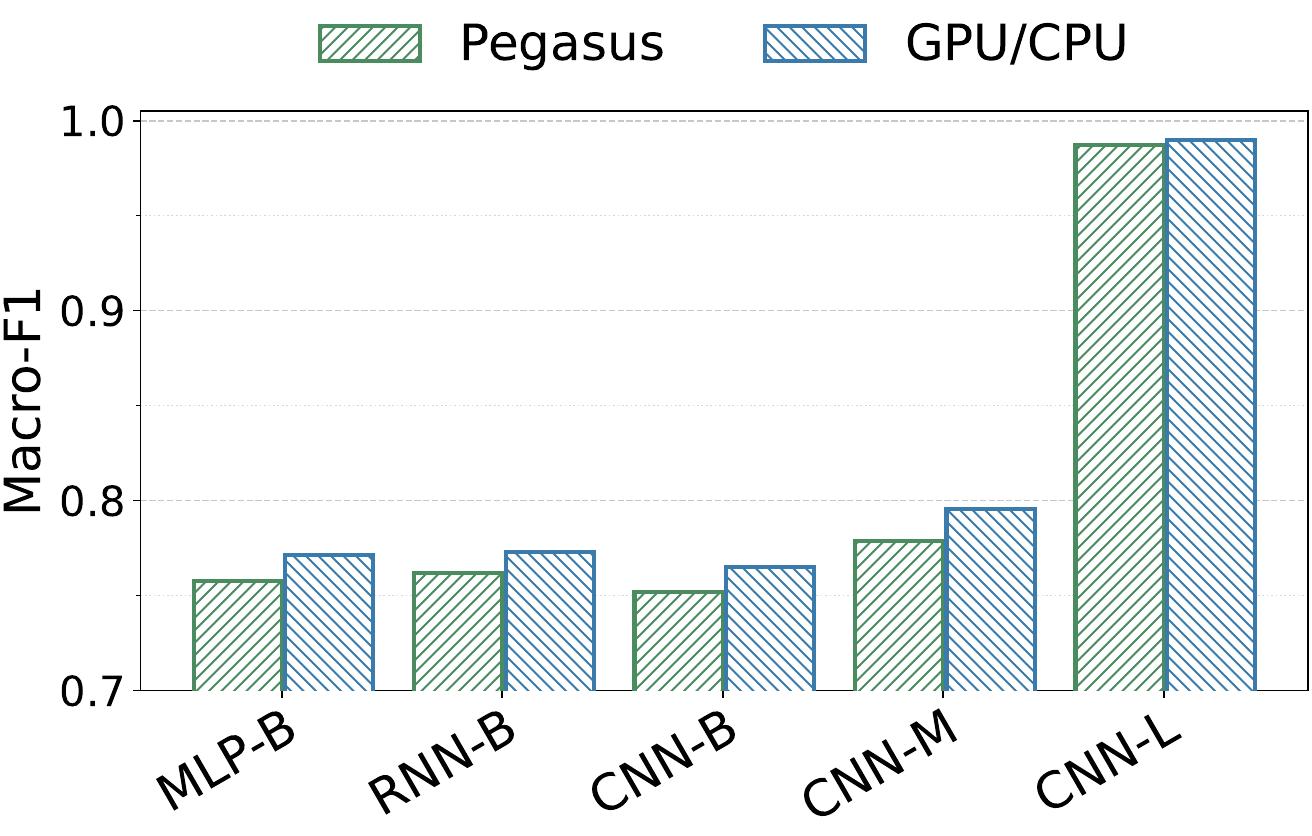}\label{fig:degration_VPN}}
    \subfigure[Throughput]{\includegraphics[width=0.24\textwidth]{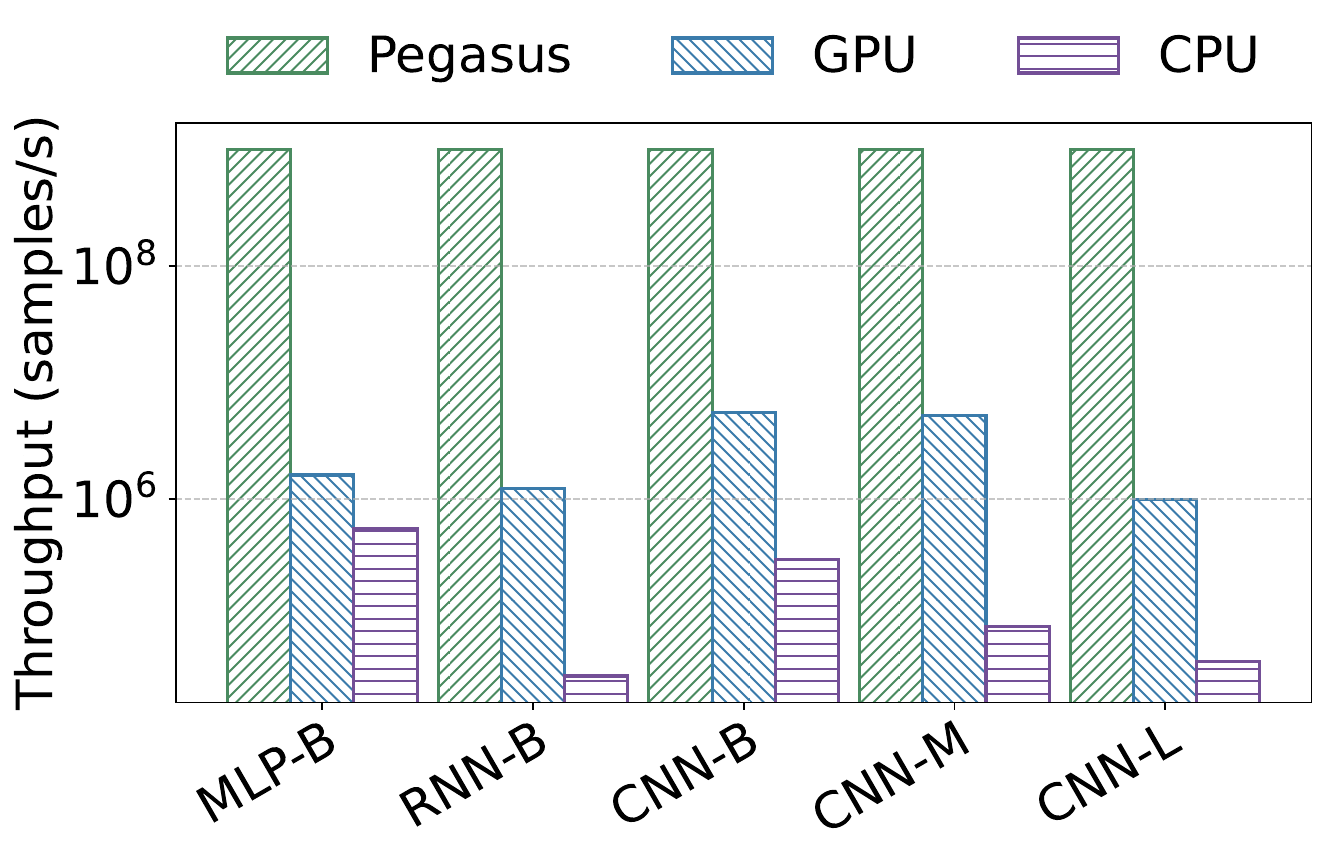}\label{fig:throughput}}
    \caption{(a–c) Comparison of classification accuracy for different models implemented on the programmable switch v.s. CPU/GPU across various datasets. (d) Throughput comparison for models on the programmable switch v.s. CPU/GPU.}
    \label{fig:degration}
\end{figure*}

\TIDP uses fuzzy matching instead of precise computation to perform DL inference, which inevitably reduces model accuracy. 
To evaluate the impact of \TIDP on accuracy and throughput improvements, we implemented full-precision DL inference on the edge using an Intel Xeon E5-2699 v4 CPU and four Tesla V100 GPUs.
Since the PISA pipeline on programmable switches ensures that any program compiled for it can run at line-rate, the size of the DL model does not affect dataplane throughput. 
To maximize control plane throughput, we pre-loaded features into CPU memory and GPU VRAM, using multi-threading to fully utilize all CPU cores and four GPUs, minimizing communication overhead. The accuracy and throughput comparison results are shown in the Figure~\ref{fig:degration}.

The results indicate that \TIDP results in an average reduction of 1.08\% in model accuracy, ranging from 0.2\% to 1.7\%. 
Notably, the CNN-L model, which features richer inputs and higher model capacity, experiences below-average accuracy loss (0.3\%, 0.2\%, and 0.9\% across three datasets). This encourages us to fully leverage \textsf{Pegasus}’s potential in designing more powerful models, rather than limiting its use to simple, small-scale neural networks.

However, it increases throughput by over $3800\times$ and $600\times$ compared to CPU and GPU, respectively. 
This throughput improvement represents the idealized capacity of these devices. 
In real-world conditions, although the gap may be narrower, the throughput gains would still be significant. Given the substantial throughput improvement, the reduction in accuracy can be considered acceptable.
\section{Discussion and Related Work}\label{discuss}

\noindent \textbf{Data-Driven Traffic Analysis.}
Researchers have proposed various methods for intelligent traffic analysis~\cite{liang2019neural}, such as encrypted traffic classification, website fingerprinting and malicious traffic detection. 
There is a growing recognition of the benefits of performing traffic analysis at line-rate. 
NetBeacon \cite{netbeacon} and Leo \cite{leo} leverage decision trees in the dataplane to enable IDP. 
However, like N3IC \cite{n3ic}, these methods face the challenge of extracting complex features directly on the dataplane. 
BoS \cite{bos} addresses this limitation by using DL to automatically extract features, supporting IDP without manual feature engineering on the dataplane. 
Building on this foundation, \TIDP further enhances the capability of executing deep learning inference within the dataplane.

\noindent \textbf{Hardware Dependency.}
Taurus \cite{taurus}, Trio \cite{yang2022trio}, Trident \cite{broadcom} have explored adding computational capabilities to the dataplane to support IDP.
However, achieving line-rate computation is often prohibitively expensive and difficult to integrate with the fundamental operations of the dataplane, such as table lookups and packet forwarding.
In contrast, \TIDP is specifically designed to align with the flow-table-centric architecture of the dataplane. Operations like multi-level comparisons and fixed-point addition, which \TIDP relies on, can be more efficiently implemented on other dataplane devices. 
In fact, the majority of our overhead is caused by the limitation of the Barefoot Tofino architecture. We believe that with lightweight hardware adjustments, \TIDP could enable more advanced capabilities for IDP.

\noindent \textbf{Deployment in Real-World Environments.}
\TIDP is designed to implement DL models on the dataplane, allowing users to balance accuracy and resource overhead based on their specific requirements. 
As such, we did not focus on the challenges of running multiple applications simultaneously on a single programmable switch in real-world deployments. Additionally, IDP may encounter issues with limited flow registers, particularly in extreme conditions. This limitation is inevitable in scenarios requiring stateful functionalities~\cite{lerner2024rethinking}.
For such challenges, prior works such as AIFO \cite{AIFO} and P4LRU \cite{zhao2023p4lru} offer more relevant solutions. These methods can store flow characteristics for large flows and utilizing packet features to identify small flows, ultimately achieving higher classification accuracy.

\section{Conclusion}\label{conc}

The limited computational resources of programmable switc-hes are not the root cause hindering intelligence realization; rather, it is the ineffective use of the MAT abstraction that becomes the real obstacle. 
Simple but useful, \TIDP expresses DL models using dataplane-friendly primitives, enabling implementation on commodity programmable switches without requiring additional complex computational resources.
The primary goal of \TIDP is to address the accuracy, scalability, and generality limitations of prior IDP designs. Experimental results demonstrate \textsf{Pegasus}’s advantages in realizing intelligent models. It serves as a viable option for line-rate DL inference and offers an alternative to the growing trend of continuously adding line-rate computational resources to the dataplane.

\noindent \textbf{Ethics:} \textit{This work does not raise any ethical issues.}
\bibliographystyle{ACM-Reference-Format}
\bibliography{mybib}


\end{document}